\documentclass[12pt]{article}    % Specifies the document style.
\usepackage[ansinew]{inputenc} % permite utilizar los acentos y la ñ con el teclado directamente
\usepackage{hyperref}
\usepackage{graphics}
\usepackage{epsfig}
\usepackage{graphicx} %for graphics inclusion
\usepackage{amssymb}  %provides various useful mathematical symbols
\usepackage{amsmath}
\usepackage{multirow} %provides the command needed for spanning rows
\usepackage{enumitem} % leftmargin - itemize
\usepackage{color}     % para colorear el texto
\usepackage{indentfirst}  % para sangran el primer párrafo de cada sección
\begin{document}
%%%%%%%%%%%%%%%%%%%%%%%%%%%%%%%%%%%%%
\begin{center}

\section*{Modelling major failures in power grids in the whole range}

\vskip 0.2in {\sc \bf Faustino Prieto$^a$
\footnote{Corresponding author. Tel.: +34 942 206758; fax: +34 942 201603. E-mail address: faustino.prieto@unican.es (F. Prieto).},
Jos\'e Mar\'{\i}a Sarabia$^a$ and Antonio Jos\'e S\'aez$^b$

\vskip 0.2in

{\small\it $^a$Department of Economics, University of Cantabria, Avenida de los Castros s/n, 39005 Santander, Spain\\
           $^b$Department of Statistics and Operations Research, Polytechnic School of Linares, University of Ja\'en, 
               C/ Alfonso X El Sabio 28, 23700 Linares (Jaén), Spain.
}\\
}

\end{center}

\begin{abstract}\noindent
Empirical research with electricity transmission networks reliability data shows that the size of major failures - in terms of energy not supplied (ENS), total loss of power (TLP) or restoration time (RT) - appear to follow a power law behaviour in the upper tail of the distribution. However, this pattern - also known as Pareto distribution - is not valid in the whole range of those major events. We aimed to find a probability distribution that we could use to model them, and hypothesized that there is a two-parameter model that fits the pattern of those data well in the entire domain.
We considered the major failures produced between 2002 and 2009 in the European power grid; analyzed those reliability indicators: ENS, TLP and RT; fitted six alternative models: Pareto II, Fisk, Lognormal, Pareto, Weibull and Gamma distributions, to the data by maximum likelihood; compared these models by the Bayesian information criterion; tested the goodness-of-fit of those models by a Kolmogorov-Smirnov test method based on bootstrap resampling; and validated them graphically by rank-size plots.
We found that Pareto II distribution is, in the case of ENS and TLP, an adequate model to describe major events reliability data of power grids in the whole range, and in the case of RT, is the best choice of the six alternative models analyzed.
\end{abstract}

\noindent {\bf Key Words}: Electricity transmission networks, Complex Systems, Power law, Pareto II, Lomax, Bootstrap.\\[1ex]
\noindent {\bf PACS numbers}: 02.70.-c, 84.70.+p, 89.75.-k.

\newpage
\section{Introduction}

Electricity transmission networks provide the means to transport the electricy from the power plants, where is produced, to the distribution networks, near our homes and businesses. Unfortunately, failures in these systems do happen - and nowadays, electricity is essential for all of us. For that reason, the analysis of those failure events, in particular from a statistical point of view, is crucial to improve the reliability of those transmission infrastructures \cite{Zio}. 

In this direction, some promising results have been obtained using network reliability data from major events: the number of customers affected by electrical blackouts in the United States between 1984 and 2002 \cite{Clauset}; the energy not supplied, the total loss of power and the restoration time in the European power grid between 2002 and 2008 \cite{Rosas}, all can be fitted by a power law distribution (also known as Pareto distribution \cite{Pareto,Arnold}) in the upper tail of the distribution.

However, this power law behaviour is not valid in the whole range of those datasets analyzed. The number of observations included in the power-law upper tail is small. As examples, only the 15\% of the major events for energy not supplied and less than 10\% of the major events for total loss of power datasets mentioned \cite{Rosas} follow that power law behaviour.

The aim of this study was to find a probability distribution that we could use to model major events reliability data of electricity transmission networks in the whole range. Our primary hipothesis was that there is a two-parameter model that fits the pattern of data well - following the principle of parsimony and admitting more than two parameters only if necessary. The rest of this paper is organized as follows: in Section \ref{methods}, we introduce the datasets analyzed and the method used; the results are presented and discussed afterwards in Section \ref{results}; finally, the conclusions are in Section \ref{conclusions}.

\section{Data and Methods}\label{methods}

We considered the network reliability data from Union for Co-ordination of Transmission of Electricity (UCTE) \cite{UCTE} - in 2007 as a reference, an association of 29 transmission system operators of 24 european countries, with an installed capacity of 640 GW, an electricity consumption of 2600 TWh, a length of high-voltage transmission lines managed of 220000 km and 500 million people served. In 2009, all UCTE operation tasks were transferred to the European Network of Transmission System Operators for Electricity (ENTSO-E) \cite{ENTSO}. Data considered correspond to a random sample of major events, between 2002 and 2009, with Energy not Supplied (ENS) given in MWh, Total Loss of Power (TLP) given in MW, and Restoration Time (RT) given in minutes, and where zero values have not been considered. This dataset was described before in \cite{Rosas}, contains 698 major events, and can be found in \cite{Data}. Table \ref{data} shows the main empirical characteristics of ENS, TLP and RT.
\begin{table}[ht]\footnotesize
  \renewcommand{\tablename}{\footnotesize{Table}}
  \renewcommand\arraystretch{1.3}
  \setlength{\tabcolsep}{2.6 mm}
  \caption{\label{data}Main empirical characteristics of ENS, TLP and RT, from major events of UCTE electricity transmission network, between 2002 and 2009.}
\begin{tabular*}{1.0\textwidth}{l c c c c c c c}
        \\[-1ex]
          \hline
                  &n    &   Mean  &   Std. Dev.   &  Skewness&   Kurtosis      & Min.  & Max.\\
     \hline
ENS (MWh)         &583  &   631.17&   7133.86     &   22.31  &     521.74      &  1    & 168000   \\     
TLP (MW)          &528  &   374.41&   1431.23     &   12.05  &     178.26      &  1    & 24120    \\
RT (minutes)      &689  &   493.36&   3290.20     &   10.88  &     134.03      &  1    & 50432    \\
     \hline
        \end{tabular*}
\end{table}

We fitted and compared six models with two parameters: the Pareto II distribution (also known as Lomax distribution) \cite{Arnold,Lomax}, the Fisk distribution (also known as Log-logistic distribution) \cite{Fisk}, the Lognormal distribution \cite{Johnson}, the Pareto (Power law) distribution, the Weibull distribution \cite{Weibull} and the Gamma distribution \cite{Fisher}.
Table \ref{distributions} shows the cumulative distribution functions $F(x)$ and the probability density functions $f(x)$ of these six distributions.

\begin{table}[ht]\footnotesize
  \renewcommand{\tablename}{\footnotesize{Table}}
  \renewcommand\arraystretch{1.7}
  \setlength{\tabcolsep}{3.2 mm}
  \caption{\label{distributions}Cumulative distribution functions and probability density functions used; $\gamma(\alpha,x/\sigma)$ represents the lower incomplete gamma function.}
		\begin{tabular*}{1.0\textwidth}{l c c c}
		        \\[-1ex]
		  \hline		  
Distribution&$F(x)$&$f(x)$\\
\hline
\\[-3ex]
Pareto II
&	
$1-\left(\displaystyle\frac{x+\sigma}{\sigma}\right)^{-\alpha}$
&	
$\displaystyle\frac{\alpha \sigma^\alpha}{(x+\sigma)^{\alpha+1}}$
&
,$\;x\geq 0$
\\[2ex]
Fisk
&	
$\displaystyle\frac{1}{1+(x/\alpha)^{-\beta}}$
&	
$\displaystyle\frac{(\beta/\alpha)(x/\alpha)^{\beta-1}}{(1+(x/\alpha)^{\beta})^2}$
&
,$\;x>0$
\\[2ex]
Lognormal
&	
$\Phi\left(\displaystyle\frac{\log x-\mu}{\sigma}\right)$
&
$\displaystyle\frac{1}{x\sigma\sqrt{2\pi}}\exp\left[-\frac{(\log x-\mu)^2}{2\sigma^2}\right]$
&
,$\;x>0$
\\[2ex]
Pareto (PowerLaw)
&	
$1-\left(\displaystyle\frac{x}{\sigma}\right)^{-\alpha}$
&	
$\displaystyle\frac{\alpha \sigma^\alpha}{x^{\alpha+1}}$
&
,$\;x\geq\sigma$
\\[2ex]
Weibull
&	
$1-\exp\left[-\left(\displaystyle\frac{x}{\lambda}\right)^\beta\right]$
&	
$\left(\displaystyle\frac{\beta}{\lambda}\right)\left(\displaystyle\frac{x}{\lambda}\right)^{\beta-1}\exp\left[-\left(\displaystyle\frac{x}{\lambda}\right)^\beta\right]$
&
,$\;x\geq 0$
\\[2ex]
Gamma
&	
$\displaystyle\frac{1}{\Gamma(\alpha)}\gamma\left(\alpha,\displaystyle\frac{x}{\sigma}\right)$
&	
$\displaystyle\frac{1}{\Gamma(\alpha)\sigma^\alpha}x^{\alpha-1}\exp\left(-\displaystyle\frac{x}{\sigma}\right)$
&
,$\;x>0$
\\[2ex]
     \hline
%     \\[0pt] % espacio entre la tabla y el texto siguiente.
		\end{tabular*}
\end{table}

First, we fitted all six models by maximum likelihood \cite{Fisher}. For each model, the log-likelihood function is given by, 
\begin{equation}\label{loglikelihood}
\log\ell(\theta|x)=\sum_{i=1}^n\log f(x_i|\theta),
\end{equation} 
where $\theta$ is the unknown parameter vector of the model, $x$ is the sample data, $f(x)$ is its probability density function showed in Table \ref{distributions}, and the maximum likelihood estimation of the parameter vector $\hat{\theta}$ is the one that maximizes the likelihood function $\log\ell(\theta|x)$. 

Then, we compared those models using the following model selection criteria: the Akaike information criterion ($AIC$), defined by  \cite{Akaike} 
\begin{equation}\label{aic}
AIC=-2\log L+2d;
\end{equation}
and the Bayesian information criterion ($BIC$), defined by \cite{Schwarz}
\begin{equation}\label{bic}
BIC=\log L-\frac{1}{2}d\log n;
\end{equation}
where $\log L=\log\ell(\hat{\theta}|x)$ is the log-likelihood (see Eq. \ref{loglikelihood}) of the model evaluated at the maximum likelihood estimates, $d$ is the number of parameters, $n$ is the number of data, and the model chosen is the one with the smallest value of $AIC$ statistic or with the largest value of $BIC$ statistic. 

After that, we tested the goodness-of-fit of all the six models considered by a Kolmogorov-Smirnov ($KS$) test method based on bootstrap resampling \cite{Clauset,Efron,Wang,Babu}. Let $x_{1},x_{2},\ldots, x_{n}$ be the sample of $X$ and $$F_n(x_{i})\approx\displaystyle\frac{1}{n+1} \displaystyle\sum_{j=1}^n I_{[x_{j}\leq x_{i}]}$$ be the empirical cumulative distribution function (cdf) in a sample value with the indicated plotting position formula \cite{Castillo}. Let $F(x;\hat{\theta})$ be the theoretical cdf of a particular model fitted by maximum likelihood. The $KS$ statistic of the model is given by \cite{Kolmogorov,Smirnov}
\begin{equation}\label{KS}
D_n= \sup\;\lvert F_n(x_{i})-F(x_{i};\hat{\theta}) \rvert,\;i=1,2,\dots,n,
\end{equation}
and the null hypothesis to test is {\it $H_0$: the data follow that model}. Then, for each model, the procedure is as follows: 
(1) calculate the empirical $KS$ statistic for the observed data; 
(2) generate, by simulation, enough synthetic data sets (in this study, we generated 10000 data sets), with the same sample size $n$ as the observed data - if $U$ is uniformly distributed on $[0,1]$ and $Q(p,\hat{\theta})$ is the theoretical quantile function of the model, then $Q(U,\hat{\theta})$ has that model distribution; 
(3) fit each synthetic data set by maximum likelihood and obtained its theoretical cdf;
(4) calculate the $KS$ statistic for each synthetic data set - with its own theoretical cdf;
(5) calculate the $p$-value as the fraction of synthetic data sets with a $KS$ statistic greater than the empirical $KS$ statistic;
(6) null hypothesis can be rejected with the 0.05 level of significance if $p$-value$<0.05$.

Finally, as a graphical model validation, we used a rank-size plot (on a log-log scale). Let $x_{(1)}\le x_{(2)}\le\dots\le x_{(n)}$ be the ordered sample of $X$, we considered the scatter plot of the points (observed data)
\begin{equation}\label{graph1}
\log[rank_i]\;\;\mbox{versus}\;\;\log[x_{(i)}],\;\;i=1,2,\dots,n,
\end{equation}
where $rank_i=n+1-i=(n+1)(1-F_n(x_{(i)})$, plotted it together with the complementary of the theoretical cdf of the model multiplied by $(n+1)$
\begin{equation}\label{graph2}
\log[(n+1)(1-F(x_{(i)};\hat{\theta})]\;\;\mbox{versus}\;\;\log[x_{(i)}],\;\;i=1,2,\dots,n,
\end{equation}
and evaluated graphically how well the model fitted the observed data.

\section{Results and Discussion}\label{results}

Tables \ref{table1},\ref{table2} show the parameter estimates and their standard errors from the six alternative models considered: the Pareto II distribution ($\alpha$ and $\sigma$ parameters); the Fisk distribution ($\beta$ and $\alpha$ parameters); the Lognormal distribution ($\mu$ and $\sigma$ parameters); the Pareto (power law) distribution ($\alpha$ and $\sigma$ parameters); the Weibull distribution ($\beta$ and $\lambda$ parameters) and the Gamma distribution ($\alpha$ and $\sigma$ parameters); fitted to the Energy not Supplied (ENS), Total Loss of Power (TLP) and Restoration Time (RT) datasets in the whole range, by maximum likelihood. 

Table \ref{table3} shows the values of $BIC$ statistic (Eq. \ref{bic}), obtained from the six candidate models, corresponding to ENS, TLP and RT datasets in the whole range. Pareto II model presents the largest value of $BIC$ in ENS and RT datasets, followed by the Fisk and Lognormal distribution. With respect to TLP dataset, Fisk, Pareto II and Lognormal models present the largest values of $BIC$ - these three results are very similar and slightly better for Fisk model. Therefore, Pareto II is the preferable model in ENS and RT datasets; and Pareto II, Fisk and Lognormal models are the preferable models in TLP dataset, according to Bayesian information criterion - denote that $AIC$ statistics (Eq. \ref{aic}) provide, in this case, equivalent results to these for the $BIC$ statistics.

\begin{table}[p]\footnotesize
  \renewcommand{\tablename}{\footnotesize{Table}}
  \renewcommand\arraystretch{1.5}
  \setlength{\tabcolsep}{2.4 mm}
  \caption{\label{table1}
Parameter estimates from the Pareto II, Fisk and Lognormal models to the ENS, TLP and RT datasets by maximum likelihood (standard errors in parenthesis).
}
  \centering
        \begin{tabular*}{1.0\textwidth}{l c c c c c c c c c}
        \\[-1ex]
          \hline
\multirow{2}{*}{Data Set} &&\multicolumn{2}{c}{Pareto II} && \multicolumn{2}{c}{Fisk}         && \multicolumn{2}{c}{Lognormal} \\
\cline{3-4}
\cline{6-7}
\cline{9-10}
                          &&$\hat{\alpha}$    &$\hat{\sigma}$&& $\hat{\beta}$    &$\hat{\alpha}$ && $\hat{\mu}$      &$\hat{\sigma}$ \\
\hline
\multirow{2}{*}{ENS}      &&   0.6445         &  10.578     &&    0.8678         &   21.787      && 3.2351           &  2.0546   \\[-1.5ex]
                          &&  (0.0428)        & (1.4033)    &&   (0.0299)        &  (1.8136)     && (0.0851)         & (0.0602)  \\[-0.5ex]
\multirow{2}{*}{TLP}      &&   1.1953         &  115.56     &&    1.0787         &   89.034      && 4.4894           &  1.6495   \\[-1.5ex]
                          &&  (0.1146)        & (18.127)    &&   (0.0390)        &  (6.2531)     && (0.0718)         & (0.0508)  \\[-0.5ex]
\multirow{2}{*}{RT}       &&   0.7768         &  17.896     &&    0.9819         &   26.210      &&  3.4172          &  1.8521   \\[-1.5ex]
                          &&  (0.0499)        & (2.1207)    &&   (0.0312)        &  (1.7666)     && (0.0706)        & (0.0499)  \\  
     \hline
        \end{tabular*}
\end{table}
\begin{table}[p]\footnotesize
  \renewcommand{\tablename}{\footnotesize{Table}}
  \renewcommand\arraystretch{1.5}
  \setlength{\tabcolsep}{2.4 mm}
  \caption{\label{table2}
Parameter estimates from the Pareto, Weibull and Gamma models to the ENS, TLP and RT datasets by maximum likelihood (standard errors in parenthesis).
}
  \centering
        \begin{tabular*}{1.0\textwidth}{l c c c c c c c c c}
        \\[-1ex]
          \hline
\multirow{2}{*}{Data Set} &&\multicolumn{2}{c}{Pareto}   && \multicolumn{2}{c}{Weibull}        && \multicolumn{2}{c}{Gamma}\\
\cline{3-4}
\cline{6-7}
\cline{9-10}
                     &&$\hat{\alpha}$    &$\hat{\sigma}$ && $\hat{\beta}$    &$\hat{\lambda}$  && $\hat{\alpha}$  &$\hat{\sigma}$\\
\hline
\multirow{2}{*}{ENS} &&   0.3091         &  1.0000      &&   0.4128         &  75.802         &&  0.2249         &   2806.5      \\[-1.5ex]
                     &&  (0.0154)        & (0.0896)     &&  (0.0113)        &  (8.0903)       && (0.0102)        &  (276.05)     \\[-0.5ex]
\multirow{2}{*}{TLP} &&   0.2227         &  1.0000      &&   0.5930         &  203.14         &&  0.4497         &   832.47      \\[-1.5ex]
                     &&  (0.0110)        & (0.1046)     &&  (0.0179)        &  (15.815)       && (0.0226)        &  (68.357)     \\[-0.5ex]
\multirow{2}{*}{RT}  &&   0.2926         &  1.0000      &&   0.4398         &  82.772         &&  0.2544         &   1939.0      \\[-1.5ex]
                     &&  (0.0133)        & (0.0837)     &&  (0.0109)        &  (7.6332)       && (0.0107)        &  (167.55)     \\  
     \hline
        \end{tabular*}
\end{table}
\begin{table}[p]\footnotesize
  \renewcommand{\tablename}{\footnotesize{Table}}
  \renewcommand\arraystretch{1.5}
  \setlength{\tabcolsep}{3.4 mm}
  \caption{\label{table3}
$BIC$ statistics for six candidate models, fitted for ENS, TLP and RT datasets in the entire domain. Larger values indicate better fitted models.} 
  \centering
        \begin{tabular*}{1.0\textwidth}{l c c c c c c c}
        \\[-1ex]
     \hline
Data Set&  Pareto II     &   Fisk              &   Lognormal          &   Pareto      &   Weibull        &  Gamma     \\
     \hline
ENS     &   -3125.2      &   -3137.4           &    -3139.5           &   -3159.9     &   -3254.3        &  -3456.6   \\[-0.5ex]
TLP     &   -3389.7      &   -3389.4           &    -3390.1           &   -3697.6     &   -3438.7        &  -3502.7    \\[-0.5ex]
RT      &   -3744.0      &   -3751.4           &    -3763.3           &   -3896.7     &   -3918.7        &  -4138.9    \\   
     \hline
        \end{tabular*}
\end{table}

Tables \ref{table4},\ref{table5} show, respectively, the values of Kolmogorov-Smirnov ($KS$) statistic (Eq. \ref{KS}) and the $p$-values obtained by bootstrap resampling, from the six alternative models analyzed, corresponding to ENS, TLP and RT datasets in the entire domain. With respect to ENS dataset, the null hypothesis {\it $H_0$} for Pareto II model cannot be rejected ($p$-value $=0.0727\geq0.05$) and for the rest five models (Fisk, Lognormal, Pareto, Weibull and Gamma) can be rejected ($p$-value $<0.05$) at the 0.05 level of significance. In the case of TLP dataset, {\it $H_0$} for Pareto II, Fisk and Log-normal models cannot be rejected and for Pareto, Weibul and Gamma can be rejected at the 0.05 level of significance. Finally, {\it $H_0$} for all the six models can be rejected at the 0.05 level of significance in the case of RT dataset.
\begin{table}[ht]\footnotesize
  \renewcommand{\tablename}{\footnotesize{Table}}
  \renewcommand\arraystretch{1.5}
  \setlength{\tabcolsep}{3.4 mm}
  \caption{\label{table4}
Empirical $KS$ statistics for the six candidate models in the entire domain of the ENS, TLP and RT datasets.} 
  \centering
        \begin{tabular*}{1.0\textwidth}{l c c c c c c c}
        \\[-1ex]
     \hline
Data Set&  Pareto II     &   Fisk               &  Lognormal           &   Pareto      &   Weibull        &  Gamma     \\
     \hline  
ENS     &   0.0323       &   0.0447             &    0.0695            &   0.1642      &   0.1150         &  0.2588    \\[-0.5ex]
TLP     &   0.0266       &   0.0240             &    0.0213            &   0.3348      &   0.0755         &  0.1481    \\[-0.5ex]
RT      &   0.0402       &   0.0522             &    0.0664            &   0.2131      &   0.1335         &  0.2692    \\   
     \hline
        \end{tabular*}
\end{table}
\begin{table}[ht]\footnotesize
  \renewcommand{\tablename}{\footnotesize{Table}}
  \renewcommand\arraystretch{1.5}
  \setlength{\tabcolsep}{3.4 mm}
  \caption{\label{table5}
Bootstrap $p$-values for the six candidate models in the entire domain of the ENS, TLP and RT datasets. Values of $p<0.05$ indicate that the models can be ruled out with the 0.05 level of significance.} 
  \centering
        \begin{tabular*}{1.0\textwidth}{l c c c c c c c}
        \\[-1ex]
     \hline
Data Set&  Pareto II     &   Fisk               &  Lognormal           &   Pareto      &   Weibull        &  Gamma     \\
     \hline  
ENS     &   0.0727       &   0.0004             &    0.0000            &   0.0000      &   0.0000         &  0.0000    \\[-0.5ex]
TLP     &   0.3640       &   0.4522             &    0.2720            &   0.0000      &   0.0000         &  0.0000    \\[-0.5ex]
RT      &   0.0013       &   0.0000             &    0.0000            &   0.0000      &   0.0000         &  0.0000    \\   
     \hline
        \end{tabular*}
\end{table}

\begin{figure}[ht]
\renewcommand{\figurename}{\footnotesize{Figure}}
\begin{center}
\includegraphics*[scale=1.0]{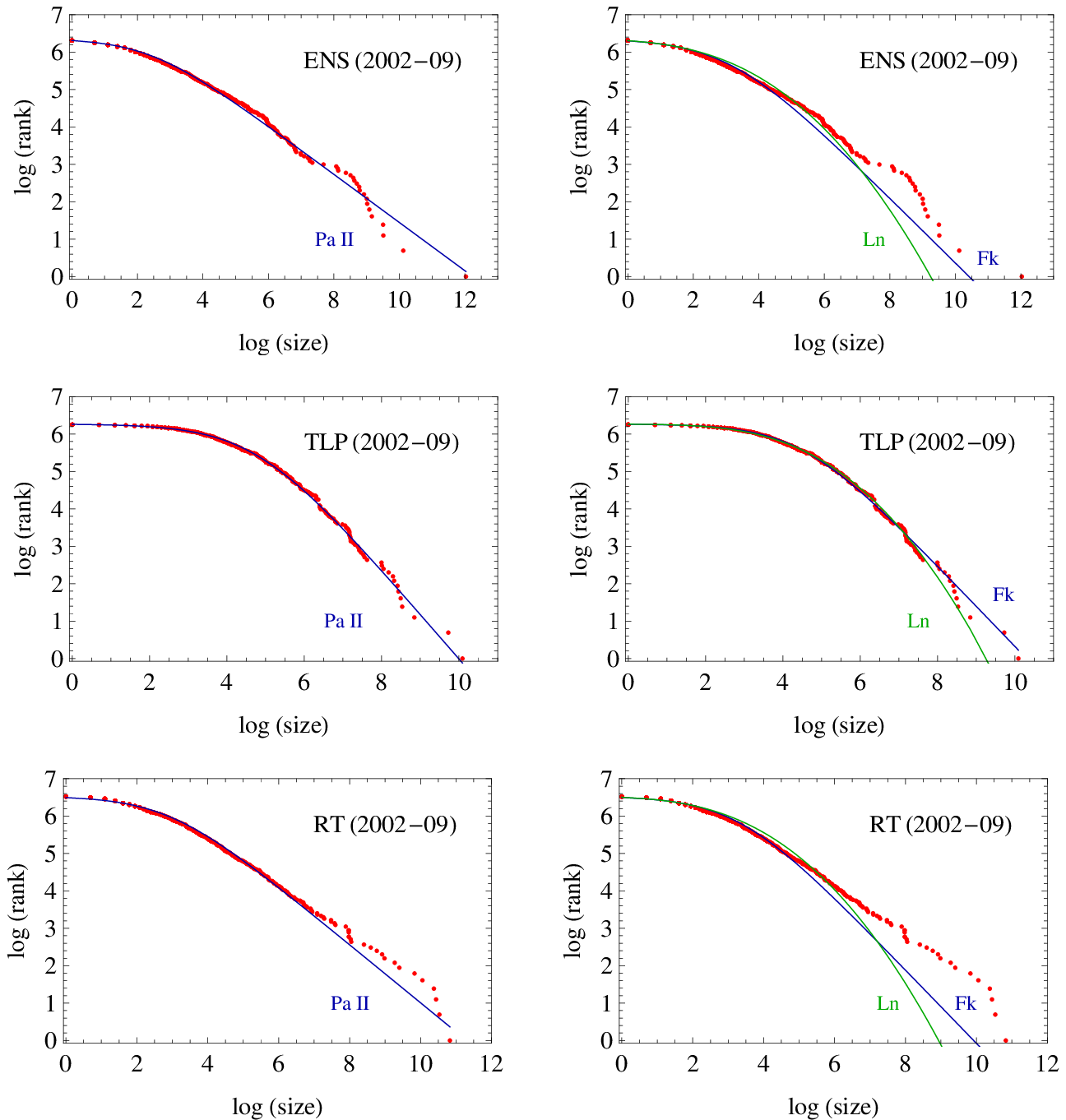}
\caption{\label{fig01}\footnotesize{Rank-size plots of the complementary of the cdf multiplied by $n+1$ (solid lines) of the Pareto II (Pa II), Fisk (Fk) and Lognormal (Ln) distributions and the observed data, on log-log scale. Left: Pareto II model. Right: Fisk and Lognormal models. Data: energy not supplied (ENS), total loss of power (TLP) and restoration time (RT), from european power grid major events in the entire domain, in the period 2002-2009.}}
\end{center}
\end{figure}

Rank-size plots (\ref{graph1},\ref{graph2}) corresponding to major events between 2002 and 2009, in the whole range, of Energy not Supplied (ENS, in MWh), Total Loss of Power (TLP, in MW) and Restoration Time (RT, in minutes) datasets, show graphically (see figure \ref{fig01}): the adequacy of the Pareto II model to the ENS dataset in contrast to the Fisk and Lognormal distributions; the adequacy of the Pareto II, Fisk and Lognormal models to the TLP dataset; and the best description of the RT dataset given by the Pareto II model in comparison with Fisk and Lognormal models. 

In summary, according to the results obtained, Pareto II distribution may serve as an adequate model for Energy Not Supplied and Total Loss of Power data from major failures in Electricity Transmission Networks in the entire domain. Adittionally, Pareto II distribution fits reasonably well Restoration Time data but with some deviation, improving other alternative models such as Fisk, Lognormal, Pareto, Weibull and Gamma distribution - unfortunately, this deviation is statistically significant.
Note that Pareto II distribution is a shifted power law distribution, which turns into a Pareto distribution for large values of the variable \cite{Milojevic}, following the known power law behaviour in the upper tail, and has only two parameters which means simplicity. For all of that, we think that Pareto II (Lomax) distribution is a good alternative for modelling power grid reliability data, in the entire domain of the major events.

\section{Conclusions}\label{conclusions}

We found a two parameter probability distribution that we can use to model major events reliability data of electricity transmission networks in the entire domain: the Pareto II distribution - also known as Lomax distribution. 

Pareto II model fits very well the pattern of Energy not Supplied (ENS) and Total Loss of Power (TLP) data and is the best of the six models considered for Restoration Time (RT) data. Additionaly, we found other two models with two parameters: the Fisk (also known as Log-logistic distribution) and the Lognormal distributions, adequate especifically for Total Loss Power data. 

We considered the major failures produced between 2002 and 2009 in the European power grid operated by UCTE; analyzed three reliability indicators: ENS, TLP and RT; fitted six alternative models: Pareto II, Fisk, Lognormal, Pareto (PowerLaw), Weibull and Gamma distributions, to the data by maximum likelihood; compared these models by the Bayesian information criterion; tested the goodness-of-fit of those models by a Kolmogorov-Smirnov test method based on bootstrap resampling; and validated them graphically by rank-size plots.

Future work is needed to find a better model for Restoration Time data from major failures in power grids in the entire domain - with two parameters or three parameters if necessary.

Previous empirical research has shown that Pareto (power law) distribution is an adequate model to describe major events reliability data of electricity transmission networks in the upper tail. In this study we found that Pareto II distribution - a shifted power law distribution - is a better choice to describe major events reliability data of electricity transmission networks in the entire domain.

\section*{Acknowledgements}

The authors thank to Ministerio de Econom\'{\i}a y Competitividad (project ECO2010-15455) for partial support of this work. We thank Mart\'{\i} Rosas Casals for his assistance with data collection. 

\bibliographystyle{plain}

\end{document}